\documentclass[twocolumn,preprintnumbers,
               superscriptaddress]{revtex4}
\usepackage{graphicx, fancybox}
\usepackage{amsmath,amssymb}

  \newcommand{\bm}[1]{\mbox{\boldmath$#1$}}

\newcommand{\Tr}{\mbox{Tr}}
\newcommand{\feyn}[1]{
  \setbox0=\hbox{\ensuremath{#1}}
  \hbox to\wd0{\hbox to0pt{\hbox to\wd0{\hss/\hss}\hss}\box0}}

\begin{document}

\title{
Bound diquarks and their Bose-Einstein condensation \\
in strongly coupled quark matter 
}

\author{Masakiyo Kitazawa}
\email{masky@yukawa.kyoto-u.ac.jp}
\affiliation{Department of Physics, Osaka University, 
Toyonaka, Osaka 560-0043, Japan}

\author{Dirk H.\ Rischke}
\email{drischke@th.physik.uni-frankfurt.de}
\affiliation{Institut f\"ur Theoretische Physik
and Frankfurt Institute for Advanced Studies, 
J.W.\ Goethe-Universit\"at,
D-60438 Frankfurt am Main, Germany}

\author{Igor A.\ Shovkovy}
\email{shovkovy@th.physik.uni-frankfurt.de}
\affiliation{
Western Illinois University, Macomb, IL, 61455 USA}

\begin{abstract}

We explore the formation of diquark molecules
and their Bose-Einstein condensation (BEC) 
in the phase diagram of three-flavor quark matter
at nonzero temperature, $T$, and quark chemical potential,
$\mu$.
Using a quark model with a four-fermion interaction, we
identify possible diquark excitations as poles of
the microscopically computed diquark propagator.
The quark masses are obtained
by solving a dynamical equation for the chiral
condensate and are found to determine the stability
of the diquark excitations. 
The stability of diquark excitations is investigated 
in the $T-\mu$ plane for different values of the diquark 
coupling strength. We find that bound diquark
molecules appear at small quark chemical potentials
at intermediate coupling and that BEC of non-strange
diquark molecules occurs if the attractive interaction 
between quarks is sufficiently strong.

\end{abstract}

\date{\today}
\maketitle


\section{Introduction}  \label{intro}

The one-gluon exchange interaction between quarks is
attractive in the color-antitriplet channel and leads
to color superconductivity in cold and dense quark matter
\cite{reviews}. Due to asymptotic freedom of QCD,
the strength of the gluon-exchange interaction around the
Fermi surface changes with the quark number density.
At asymptotically high densities,
the interaction is sufficiently weak to apply perturbation
theory and the mean-field approximation, and 
color superconductivity is very similar to standard
BCS superconductivity. The gap parameter is $\Delta \sim
\mu \exp(-1/g)$, where $g$ is the QCD coupling
constant. Quark Cooper pairs are large, with
a correlation length $\xi \sim 1/\Delta$ which is
parametrically larger than the interparticle
distance $d \sim 1/\mu$. On the other hand, 
at intermediate densities which may be realized in the 
cores of compact stars and/or in the intermediate stages
of heavy-ion collisions, the quark-quark interaction 
is relatively strong and the properties of
the quark Cooper pairs will be modified. In particular, we
expect their size to become of the same order as
the interparticle distance, $\xi \sim 1/\Delta \sim d$.
Moreover, it was argued that, due to the
strong coupling, the fluctuations of 
the diquark-pair field become large 
around the critical temperature $T_c$ \cite{KKKN02} 
and that they give rise to a pseudogap region 
in the normal phase above $T_c$ \cite{Babaev:1999iz,KKKN04}.

If the quark-quark interaction becomes strong enough
before the quarks are confined at lower density,
quark Cooper pairs become small enough to be considered
as diquark molecules which are tightly bound states of two
quarks. The color-superconducting ground state of 
quark matter at low temperature turns into
a Bose-Einstein condensed phase of diquark molecules
\cite{Mat00,AHI02}.
Most likely, the transition happens continuously with
the change of the density, just like
the BCS-BEC crossover in conventional condensed matter systems
\cite{Legg80,NSR85},
see also Refs.~\cite{Nishida:2005ds,Nawa:2005sb,Abuki:2006dv,
Deng:2006ed,Ebert:2006tc,Sun:2007fc,He:2007yj}.
An interesting property of diquark molecules is that
they can exist even above the critical temperature $T_c$
of BEC, up to the dissociation temperature $T_{\rm diss} > T_c$ 
\cite{NSR85}. If such modes exist in quark matter, 
they will affect its properties and experimental observables.

In this work, we explore the appearance of diquark molecules
and their BEC in the phase diagram of quark matter
using a low-energy effective model.
This model features an attractive quark-quark interaction
with a constant coupling strength $G_D$
that is regarded as a free parameter of the model.
We show that diquark molecules appear at low density
at intermediate values of $G_D$.
It is also shown that BEC of diquark molecules can occur
for large values of $G_D$.

In the normal phase above $T_c$, the strongest decay mode 
of diquarks is that into two quarks.
Since the excitation energy of a quark at rest is $M-\mu$,
with $M$ being the mass of the quark, the threshold 
energy for this decay process
is $\omega_{\rm thr} = 2( \bar{M} - \bar\mu )$, where
$\bar{M}$ and $\bar\mu$ are the average mass
and chemical potential of the quarks in the diquark.
Recalling that the energy of the diquark excitations 
should be positive to ensure the stability of the system,
one finds that the necessary condition for the existence of 
stable diquarks is $\omega_{\rm thr} >0$, or
\begin{align}
\bar{M} > \bar\mu.
\label{eq:M>mu}
\end{align}
As we will see later, at $T=T_c$ Eq.~(\ref{eq:M>mu}) 
is also a sufficient condition \cite{NSR85,Nishida:2005ds}.
{}From Eq.~(\ref{eq:M>mu}), we conclude that the stability of 
diquark excitations is determined by the quark masses.
Quark masses are dynamically generated by 
chiral symmetry breaking
and change as functions of $T$ and $\mu$.
In this work, we solve the gap equations for 
the chiral condensates and incorporate this effect
in our calculation.
Equation~(\ref{eq:M>mu}) also indicates that 
diquarks composed of heavier quarks tend to be 
more stable than those composed of light quarks, 
provided they exist at all.
One thus expects that diquarks including 
a strange quark are more stable than those composed of
up and down quarks.

Usually, BEC is discussed using the canonical ensemble, i.e.,
the particle number density is fixed as an external parameter.
In this work, however, we employ the grand canonical 
ensemble and draw the phase diagram in the $T-\mu$ plane,
as is usually done in the literature when exploring the QCD 
phase diagram \cite{reviews}. In order to decide whether 
BEC of diquarks occurs in this ensemble,
we regard the region of the superconducting phase satisfying
Eq.~(\ref{eq:M>mu}) as Bose-Einstein condensed phase 
\cite{NSR85}.

In this exploratory study, we employ a common chemical potential 
$\mu$ for all flavors and colors. For quark matter in compact stars, 
this is probably not a very good assumption, as the chemical 
potentials should be determined to satisfy the neutrality and 
beta-equilibrium conditions. It is known that a rich phase structure 
can appear under these conditions \cite{gSC}. However, 
as we shall see in the following, diquark excitations 
play an important role even at high
temperatures and small chemical potentials in the range
relevant for heavy-ion collisions. In this case,
our assumption of equal chemical potentials for
all quark flavors and colors is applicable
to very good approximation. 

This statement warrants a few remarks.
Thermal model fits of hadron yields at chemical freeze-out
show that, because
of strangeness and isospin conservation, neither
the strangeness chemical potential $\mu_S$ nor the
isospin chemical potential $\mu_I$ are zero. The
strangeness chemical potential is nonzero 
because of associated production channels
in hadronic matter at nonzero baryon chemical potential.
For quark matter, however, $\mu_S$ should be strictly zero
if the system has zero strangeness, and we may neglect
$\mu_S$. The isospin
chemical potential is nonzero 
because of the initial isospin asymmetry of the colliding
nuclei. However, at all collision energies it has been
demonstrated \cite{BraunMunzinger:1999qy}
that $\mu_B \equiv 3 \mu \gg |\mu_I|$. 
Thus, to leading order we may set $\mu_I = 0$.

\section{Formalism} \label{form}

In order to study the phase diagram of quark matter,
we employ a three-flavor quark model with four-fermion 
interactions. The Lagrangian is given by
\begin{align}
\mathcal{L} =& \bar \psi \, ( i \partial \hspace{-0.5em} / 
- \hat{m} \, ) \psi 
+G_S \sum_{a=0}^8 \left[ \left( \bar \psi \lambda_a 
\psi \right)^2 
+ \left( \bar \psi i \gamma_5 \lambda_a \psi \right)^2 \right] 
\nonumber \\
+& G_D \sum_{\gamma,c} \left[\bar{\psi}_{\alpha}^{a} i \gamma_5
\epsilon^{\alpha \beta \gamma}
\epsilon_{abc} (\psi_C)_{\beta}^{b} \right] \left[ 
(\bar{\psi}_C)_{\rho}^{r} i \gamma_5
\epsilon^{\rho \sigma \gamma} \epsilon_{rsc} \psi_{\sigma}^{s} 
\right] ,
\label{eq:Lagrangian}
\end{align}
where the quark field $\psi_{\alpha}^{a}$ has color 
($a=r,g,b$) and flavor ($\alpha=u,d,s$) indices. 
The matrix of current quark masses is given by 
$\hat{m} = {\rm diag}_{f}(m_u, m_d, m_s)$;
$\lambda_0=\sqrt{2/3}\,\openone$ and
$\lambda_a,\, a=1, \ldots, 8,$ are
the Gell-Mann matrices in flavor space.
The charge conjugate spinors are 
$\psi_C = C \bar \psi^T$ and $\bar 
\psi_C = \psi^T C$, where $C=i\gamma^2 \gamma^0$ 
is the charge conjugation matrix.
In the following, we only consider diquark condensates
and diquark excitations in the color anti-triplet channel.
For the numerical calculations, 
we employ a three-dimensional momentum cutoff $\Lambda$. 
We treat the diquark coupling constant $G_D$ 
as a free parameter. For the other parameters,
we use the values of Ref.~\cite{Buballa:2001gj},
$m_u = m_d = 5$ MeV, $m_s=120$ MeV, $G_S=6.41\,\mbox{GeV}^{-2}$
and $\Lambda=600$ MeV.

We evaluate the thermodynamic potential in the mean-field 
approximation;
\begin{align}
  \Omega &= \frac1{4G_D}\sum_{c=1}^{3} |\Delta_c|^2+
  \frac1{8G_S} \sum_{\alpha=1}^{3}(M_\alpha-m_\alpha)^2 
\nonumber \\
  &-\frac{T}2 \sum_n \int\frac{d^3\bm{p}}{(2\pi)^3}
    \Tr_{\rm D,f,c}
  \ln \left[S^{-1}(i\omega_n,\bm{p})\right], \label{eq:Omega}
\end{align}
where the trace is taken over Dirac, flavor, and color indices,
$\omega_n = (2n+1)\pi/T$ is the Matsubara frequency for fermions, 
and
\begin{align}
M_\alpha &= m_\alpha - 4G_S \langle \bar\psi_\alpha 
\psi_\alpha \rangle,
\label{eq:M} \\
\Delta_c&= 2G_D \langle \bar\psi_C P_c \psi \rangle,
\label{eq:Delta}
\end{align}
are the constituent quark masses and the gap parameters for 
color superconductivity, respectively, 
with $(P_c)^{ab}_{\alpha\beta}
 = i\gamma_5 \epsilon^{\alpha\beta c} \epsilon_{ abc }$.
The $72\times 72$ Nambu-Gor'kov propagator is defined by
\begin{align}
   S^{-1}(i\omega_n,p)&= \left( \begin{array}{cc}
  \feyn{p}+\mu\gamma_0-\hat{M}& \sum_\eta
   P_\eta\Delta_\eta  \\
  \sum_\eta \gamma^0 P_\eta^\dagger \gamma^0 \Delta_\eta & 
  {^t\feyn{p}-\mu}\gamma_0+\hat{M}
  \end{array} \right),
\end{align}
with $\feyn{p}=i\omega_n\gamma_0-\bm{p}\cdot\bm{\gamma}$ and
$\hat{M} = {\rm diag}_{f}(M_u, M_d, M_s)$.
The quark chemical potential $\mu$ has a common value
for all flavors and colors, as mentioned above.

The physical values of the variational parameters 
$\Delta_c$ and $M_\alpha$
satisfy the stationary conditions (the gap equations) 
\begin{align}
\frac{ \partial\Omega }{ \partial \Delta_c } = 0 \quad {\rm and} \quad
\frac{ \partial\Omega }{ \partial M_\alpha } = 0.
\end{align}
As we will see later, the color-superconducting phase transitions
at nonzero temperature are of second order.
Thus, the critical temperatures are
determined by solving the following equation:
\begin{align}
\left.
\frac 1{\Delta_c} \frac{ \partial\Omega }{ \partial\Delta_c }
\right|_{\Delta_c=0} = 0.
\label{eq:crit}
\end{align}
We shall see later that this equation determines the
poles of the diquark propagator at vanishing energy and
momentum.

Since up and down flavors are degenerate
in our model, we always have
$M_u=M_d$ and $\Delta_1=\Delta_2$.
In the following, we refer to the phase with 
$\Delta_3 \ne 0$ and $\Delta_{1,2}=0$ as the 2SC phase;
the phase with $\Delta_3 \ne 0$ and $\Delta_{1,2} \ne 0$ 
is the CFL phase \cite{reviews}.
Unpaired quark matter has 
$\Delta_1=\Delta_2=\Delta_3 = 0$.
Because of the explicit chiral symmetry breaking by the
nonzero current quark masses, the chiral condensates 
$\langle \bar\psi_i \psi_i \rangle$ never vanish.

At nonzero temperature, the order parameter fields, 
$\Delta_c(\bm{x},t)$ and $M_\alpha(\bm{x},t)$ have fluctuations 
around the values determined by the mean-field approximation.
In the following, we consider the amplitude fluctuations 
of $\Delta_c(\bm{x},t)$ in unpaired quark matter.
The propagation of the fluctuations 
is characterized by the retarded propagator
\begin{widetext}
\begin{align}
D_c^R ( \bm{x},t ; \bm{x}',t' ) &= -i \theta(t-t')\langle 
[\bar{\psi}( \bm{x},t ) P_c \psi_C( \bm{x},t ), 
\bar{\psi}_C ( \bm{x}',t' ) P_c \psi(\bm{x}',t')]
\rangle 
= \int \frac{d^3 \bm{k}d\omega}{(2\pi)^4} D_c^R ( \bm{p},\omega )
{\rm e}^{ -i\omega (t-t') }
{\rm e}^{i{\bf p} \cdot ({\bf x}-{\bf x} ')},
\label{eq:D^R}
\end{align}
where $c=1,2$, and $3$ correspond to the 
down-strange, up-strange,
and up-down diquark fields, respectively.
In the random-phase approximation,
the diquark propagators are given by
\begin{align}
D^R_c(\bm{p},\omega)
= \frac{1}{2}
\frac{ Q^R_c(\bm{p},\omega) }{ 1+G_D Q^R_c(\bm{p},\omega) },
\end{align}
where $Q^R_c(\bm{p},\omega)$ is the one-loop quark-quark 
polarization function.
In the imaginary-time formalism, it is given by
\begin{equation}
\mathcal{Q}_c(\bm{p},i\nu_n) = -2T \sum_m \int 
\frac{d^3 \bm{q}}{(2\pi)^3}
\Tr_{\rm D,f,c} [P_c S_0( \bm{p}-\bm{q} , i\nu_n-i\omega_m ) 
P_c C S_0^T ( \bm{q} , i\omega_m ) C ],
\label{eq:Q}
\end{equation}
where $\nu_n = 2\pi n /T$ denotes the Matsubara frequency for bosons,
and $S_0 (\bm{p},i\omega_n) =
[ \feyn{p} +\mu \gamma_0 - \hat{M} ]^{-1}$.
Taking the analytic continuation, 
$Q_c^R(\bm{p},\omega)={\cal Q}_c(\bm{p},i\nu_n)
|_{i\nu_n\to \omega+i\eta}$,
we obtain
\begin{align}
Q^R_c(\bm{p},\omega)
&=
-2 \sum_{\beta,\gamma=1}^3 |\epsilon^{c\beta\gamma}|
\int \frac{ d^3 \bm{q} }{ (2\pi)^3 } \sum_{s,t=\pm}
\frac{ ( sE_\beta + tE_\gamma )^2 - |\bm{p}|^2 
- (\delta M_c)^2 }{ st E_\beta E_\gamma }
\frac{ f( tE_\gamma-\mu ) - f( -sE_\beta+\mu ) }
{ \omega + 2\mu - sE_\beta - tE_\gamma + i\eta },
\end{align}
where $E_\beta = \sqrt{ |\bm{q}-\bm{p}|^2 + M_\beta^2 }$, 
$E_\gamma = \sqrt{ |\bm{q}|^2 + M_\gamma^2 }$,
$\delta M_c = | M_\beta - M_\gamma |$, and 
$ f(E) = [ \exp(E/T) + 1 ]^{-1}$ is the Fermi 
distribution function. The imaginary part of 
$Q^R_c(\bm{p},\omega)$ denotes the difference of 
decay and production rates of the diquark field.
At $\bm{p}=0$, it is given by
\begin{align}
\lefteqn{ {\rm Im} Q_c^R ( \bm{0},\omega )
 = 2\pi \sum_{\beta,\gamma=1}^3 |\epsilon^{c\beta\gamma}| 
\int \frac{ d^3\bm{q} }{ (2\pi)^3 }
\frac{ ( \omega+2\mu )^2 - (\delta M_c)^2 }{ E_\beta E_\gamma } }
\nonumber \\
&\times \left\{ 
-\left[ ( 1-f_\beta^+ )( 1-f_\gamma^+ ) - f_\beta^+ 
f_\gamma^+ \right]
\delta( \omega + 2\mu - E_\beta - E_\gamma )
 +\left[ ( 1-f_\beta^- )( 1-f_\gamma^- ) - f_\beta^- 
f_\gamma^- \right]
\delta( \omega + 2\mu + E_\beta + E_\gamma ) \right.
\nonumber \\ 
& \;\;\;\;\;\, \left. -\left[ f_\beta^- ( 1-f_\gamma^+ ) - ( 1 - f_\beta^- ) 
f_\gamma^+ \right]
\delta( \omega + 2\mu + E_\beta - E_\gamma )
 -\left[ f_\beta^+ ( 1-f_\gamma^- ) - ( 1 - f_\beta^+ ) 
f_\gamma^- \right]
\delta( \omega + 2\mu - E_\beta + E_\gamma ) \right\},
\label{eq:ImQR_k0}
\end{align}
\end{widetext}
where $f_\alpha^\pm = \{ \exp[  ( E_\alpha \mp \mu )/T ] + 1 
\}^{-1}$.
The first (second) term in the bracket in Eq.~(\ref{eq:ImQR_k0})
includes the decay processes of the diquark into two quarks 
(anti-quarks) and takes nonzero values 
at $\omega > 2\bar{M}_c-2\mu$ ($\omega < -2\bar{M}_c-2\mu$),
with $\bar{M}_c = ( M_\beta + M_\gamma ) / 2 $.
The third and fourth terms represent Landau damping of
the diquark. These terms become nonzero at 
$ -\delta M_c -2\mu < \omega < \delta M_c-2\mu $.

The poles of the diquark propagator $D_c^R$ are 
determined by solving
${D_c^R({\bf p},\omega)}^{-1}=0$, or equivalently
\begin{align}
1 + G_D Q^R_c(\bm{p},\omega) =0,
\label{eq:pole}
\end{align}
in the complex-energy plane.
Applying $\omega=|{\bf p}|=0$ to this equation,
one can easily show that Eq.~(\ref{eq:pole}) is equivalent
to the critical condition Eq.~(\ref{eq:crit}).
Therefore, $D_c^R({\bf p},\omega)$ has a pole
at the origin at $T=T_c$ of the second order transition.
This property is known as the Thouless criterion 
in condensed matter physics \cite{Thou}.
Above $T_c$, the pole moves continuously  from the origin to 
the fourth quadrant. This mode is called the soft mode.
If $\bar{M}_c<\mu$ at $T=T_c$, $\omega=0$ is 
in the continuum of the decay process into two quarks and the 
imaginary part at the pole starts growing just 
above $T_c$ \cite{KKKN02}. 
When $\bar{M}_c>\mu$, on the other hand, 
the soft mode is stable against spontaneous breaking 
into a pair of quarks
and the pole moves on the real axis in the vicinity of $T_c$. 
This mode is nothing but a bound state of two quarks:
the diquark molecule \cite{NSR85}.
As $T$ increases, the pole eventually arrives at
the threshold of the decay process into two quarks
$\omega_{\rm thr}=2(\bar{M}_c-\mu)$ at the dissociation 
temperature $T_{\rm diss}$, and the soft mode is no longer 
a bound state at $T>T^c_{\rm diss}$.
Since the pole is at $\omega_{\rm thr}$ at $T^c_{\rm diss}$,
the dissociation temperature is determined by solving
\begin{align}
1 + G_D Q_c ( \bm{p}=0 , \omega_{\rm thr})
|_{T=T^c_{\rm diss}} = 0.
\end{align}

Although the diquark modes can acquire decay rates due to 
Landau damping, i.e., the third and fourth term in 
Eq.~(\ref{eq:ImQR_k0}),
our numerical results show that the soft modes never appear 
in the range of energies where Landau damping is
nonzero. Therefore, these processes do not contribute to the
decay rate of the soft modes in the parameter range 
employed in the present study.
There can appear another pole of $D^R_c(\bm{p},\omega)$
instead of the soft mode at the energy 
$-2\bar{M}_c -2\mu < \omega < -\delta M_c-2\mu$.
This mode does not have a decay rate and should be identified 
as a bound anti-diquark \cite{Nishida:2005ds}.
For lower $\mu$, thermal excitations of bound anti-diquarks 
play an important role.

If bound diquarks are formed at $T=T_c$, 
it is natural to identify the color-superconducting phase 
below $T_c$ as a Bose-Einstein condensed phase of diquark
molecules \cite{NSR85}.
In the following, therefore, we regard the color-superconducting
phase satisfying $\bar{M}_c<\mu$ as a Bose-Einstein condensate.
Notice, however, that this is just a rough guide to separate
the BEC and BCS regions; these two limits are connected 
continuously and there is no sharp phase boundary
between them \cite{NSR85}.

\section{Numerical Results}

\begin{figure}
\begin{center}
\vspace{2mm}
\includegraphics[width=.49\textwidth]{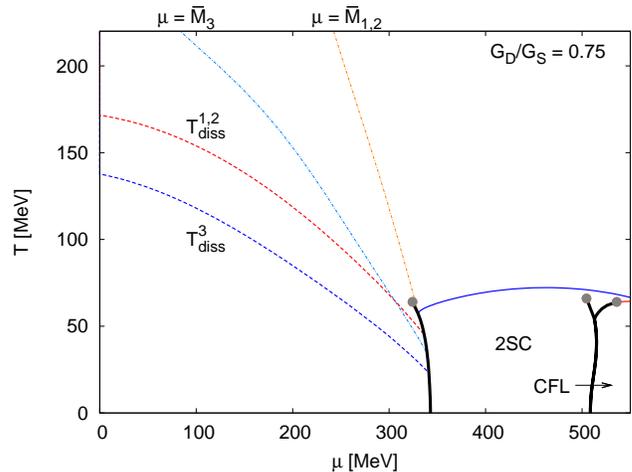} 
\caption{
The phase diagram in the $T$-$\mu$ plane for $G_D/G_S=0.75$.
The bold and thin solid lines represent first- and 
second-order phase transitions.
The dashed lines denote the dissociation temperature of 
diquark molecules for up-down diquarks, $T^3_{\rm diss}$, and 
for up-strange and down-strange diquarks, $T^{1,2}_{\rm diss}$.
The conditions $\mu=\bar{M}_c$ are 
shown by the dash-dotted lines.
}
\label{fig:phased1}
\end{center} 
\end{figure}

In this section, we show the phase diagram in the $T$-$\mu$ plane
for several values of the diquark coupling $G_D$.
In Fig.~\ref{fig:phased1},
we first discuss the case $G_D/G_S = 0.75$,
which is a value commonly used in the literature
\cite{reviews,Buballa:2001gj}.
The bold and thin solid lines represent first- and second-order
phase transitions, respectively.
One sees that there appear two types of 
color-superconducting phases, the 2SC and CFL phase, 
at high $\mu$ and low $T$.
At $T=0$, these phases are separated by a first-order 
phase transition;
the first-order transition terminates at a critical point
for some nonzero value of temperature.
The dissociation temperatures of diquark molecules
$T^c_{\rm diss}$ are shown by the dashed lines.
We see that there exists a region at small chemical potential
where stable diquark molecules are formed.

\begin{figure}
\begin{center}
\includegraphics[width=.49\textwidth]{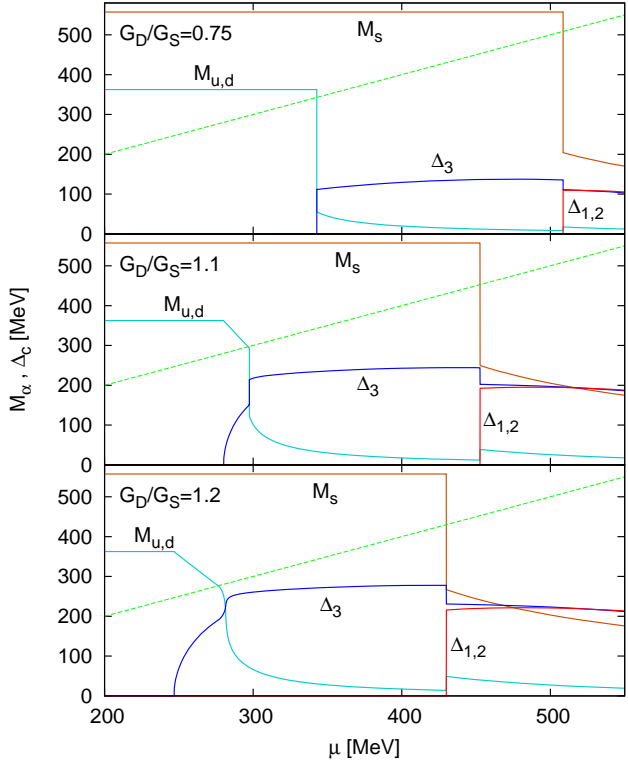} 
\caption{
Order parameters $M_\alpha$ and $\Delta_c$
at $T=0$ as functions of $\mu$ for various values of 
the diquark coupling $G_D/G_S=0.75,1.1$ and $1.2$.
The chemical potential is shown by the dashed line.
}
\label{fig:orderp}
\end{center} 
\end{figure}

In order to see whether diquark molecules undergo BEC,
we plot the conditions $\mu=\bar{M}_c$
by the dash-dotted lines in Fig.~\ref{fig:phased1}; 
the regions to the left of these lines satisfy $\mu<\bar{M}_c$.
We see that these lines terminate at the first-order transition
and we do not have a color-superconducting phase
satisfying $\mu<\bar{M}_c$.
In other words, BEC does not appear for this value
of $G_D/G_S$.
The behavior of $\bar{M}_c$ and $\Delta_c$ as functions
of $\mu$ at $T=0$ are shown in the upper panel of Fig.~\ref{fig:orderp}.
One observes that $\bar{M}_3 = M_{u,d}$ has a discontinuity at
$\mu\simeq343$ MeV corresponding to a first-order transition,
and $\bar{M}_3$ is larger than $\mu$
to the left of the discontinuity.
The diquark condensate $\Delta_3$ assumes
nonzero values only for $\mu>\bar{M}_3$.
This is a typical property at weak coupling;
when $\mu>\bar{M}_3$, the Fermi surfaces of up and
down quarks exist and the Cooper instability 
leads to a diquark condensate, while if not, 
the ground state is nothing other than the vacuum.

It is worth mentioning that bound diquark molecules
appear in the phase diagram even though 
BEC does not exist in the phase diagram.
The diquark coupling used in Fig.~\ref{fig:phased1} is
strong enough to form bound diquarks, but it is still
too weak to lead to their BEC.

\begin{figure}
\begin{center}
\vspace{2mm}
\includegraphics[width=.49\textwidth]{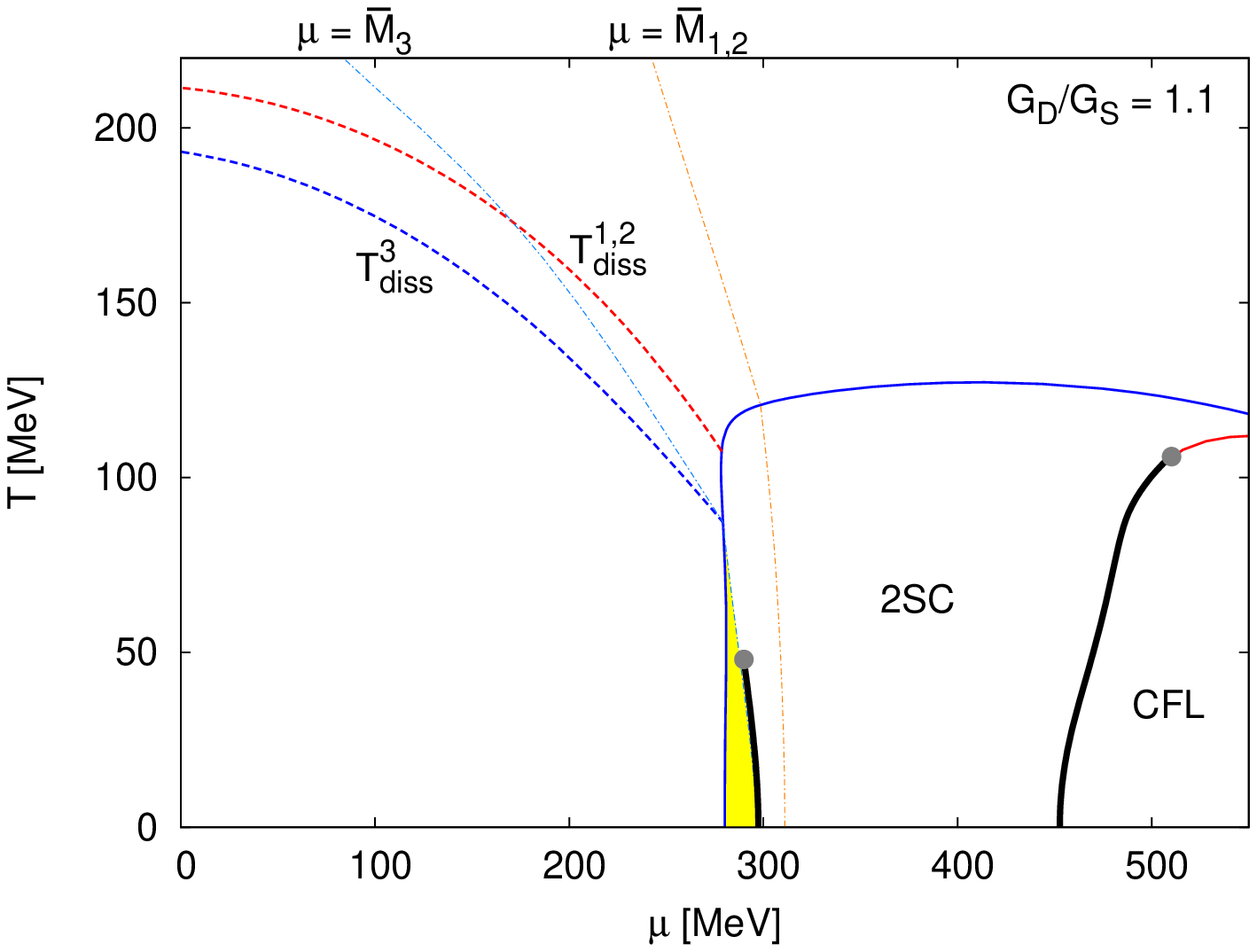} 
\\ 
\vspace{2mm}
\includegraphics[width=.49\textwidth]{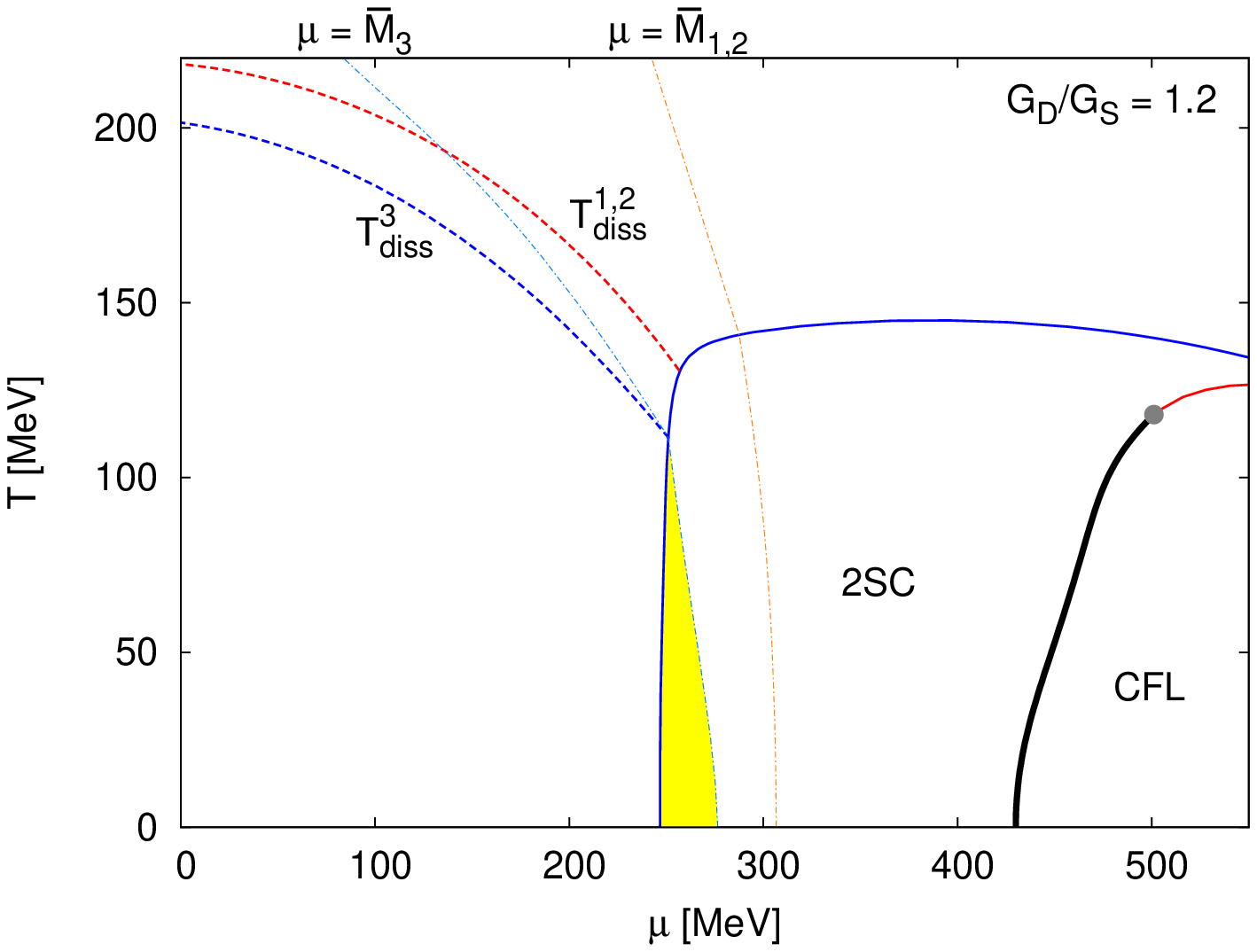} 
\caption{
The phase diagram in the $T$-$\mu$ plane
for relatively strong diquark couplings $G_D/G_S=1.1$ and $1.2$.
BEC of up-down diquarks occurs in the shaded area.
}
\label{fig:phased2}
\end{center} 
\end{figure}

Next, we show the phase diagrams with much stronger 
diquark couplings.
In Fig.~\ref{fig:phased2}, the phase diagrams
with $G_D/G_S=1.1$ and $1.2$ are shown.
We see that,  as $G_D$ becomes larger, 
the regions of the 2SC and CFL phases expand 
toward lower $\mu$ and higher $T$.
For $G_D/G_S=1.1$, there appears BEC of up-down diquarks in
the region of the 2SC phase satisfying $\mu<\bar{M}_3$,
shown by the shaded area in Fig.~\ref{fig:phased2}.
The BEC region becomes wider for $G_D/G_S=1.2$.
One also observes that the dissociation 
temperatures $T^c_{\rm diss}$ become higher as $G_D$ increases.
In the phase diagrams in Fig.~\ref{fig:phased2},
$T^c_{\rm diss}$ at $\mu=0$ are comparable or much higher than
the critical temperature of the QCD phase transition,
which is predicted to be in the range $T_c\sim 150 - 190$ MeV 
in lattice QCD simulations \cite{Lat_Tc}.
This result shows that diquark molecules can exist even
in the quark-gluon plasma phase if the diquark coupling is 
strong enough.

\begin{figure}
\begin{center}
\includegraphics[width=.45\textwidth]{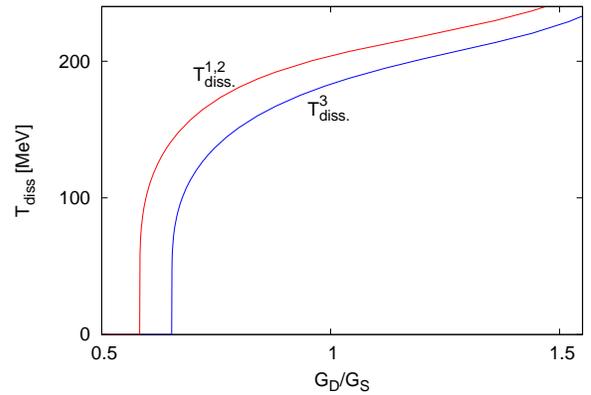} 
\caption{
Dissociation temperatures of up-down diquarks $T^3_{\rm diss}$,
and up-strange and down-strange diquarks $T^{1,2}_{\rm diss}$.
}
\label{fig:dissoc}
\end{center} 
\end{figure}

In order to see the diquark coupling dependence 
of the dissociation temperatures, we show 
$T^c_{\rm diss}$ at $\mu=0$ 
as functions of $G_D/G_S$ in Fig.~\ref{fig:dissoc}.
At weak coupling, $T^c_{\rm diss}=0$ and bound diquarks do not exist.
As $G_D/G_S$ becomes larger, $T^c_{\rm diss}$ eventually 
become nonzero and increase rapidly.
The dissociation temperatures for diquarks including 
the strange quark, $T^{1,2}_{\rm diss}$, are always higher than 
that for the up-down diquark. 
This feature comes from the difference of the threshold energy 
$2(\bar{M}_c-\mu)$.

The other interesting feature shown in 
Figs.~\ref{fig:phased1} and \ref{fig:phased2}
is the behavior of the line of first-order phase transitions.
The first-order phase transition 
at lower density is shorter for $G_D/G_S=1.1$ 
than that for $G_D/G_S=0.75$, and disappears at $G_D/G_S=1.2$. 
To understand this behavior,
we display the order parameters 
for $G_D/G_S=1.1$ and $1.2$ in the middle and lower panels of 
Fig.~\ref{fig:orderp}, respectively.
The figure shows that, as $G_D$ becomes larger, 
$\Delta_c$ increase while the quark masses $\bar{M}_c$ 
become smaller, and the discontinuity of $\bar{M}_3$ 
disappears at $G_D/G_S=1.2$.
The decrease of $\bar{M}_c$ can be understood 
as the interplay between the chiral and diquark condensates 
\cite{Kitazawa:2002bc}:
the energy gain due to diquark condensation is proportional
to the surface area of the Fermi sphere. Since the
radius of the Fermi sphere is $p_F=\sqrt{ \mu^2 - M^2}$,
the condensation energy increases when $M$ decreases.
Since the masses of quarks are suppressed as $G_D$ becomes
larger, the lines for $\mu=\bar{M}_c$ and the region of BEC 
move toward lower $\mu$.

\begin{figure}
\begin{center}
\vspace{2mm}
\includegraphics[width=.49\textwidth]{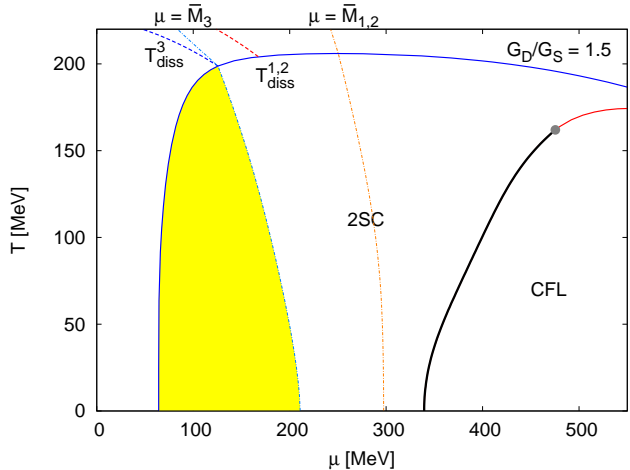} 
\caption{
The phase diagram in the $T$-$\mu$ plane
for extremely strong diquark coupling $G_D/G_S=1.5$.
}
\label{fig:phased3}
\end{center} 
\end{figure}

Finally, let us consider the phase diagram for
extremely large diquark coupling.
In Fig.~\ref{fig:phased3}, we show the phase diagram
for $G_D/G_S=1.5$.
The 2SC and CFL phases become much wider toward lower $\mu$
and higher $T$, and $T_{\rm diss}^c$ becomes higher.
The region of BEC of up-down diquarks also becomes wide.
We do not obtain a region in the CFL phase satisfying
$\mu>\bar{M}_{1,2}$, i.e., BEC of up-strange and down-strange
diquarks does not occur for $G_D/G_S \leq 1.5$. 
If the diquark coupling is raised further,
the vacuum, i.e., $T=\mu=0$, eventually becomes a Bose-Einstein
condensate of diquark molecules, which is clearly unphysical.

\section{Summary and Discussions} \label{summary}

In this paper, we explored the phase diagram of three-flavor quark 
matter focusing on the appearance of diquark molecules and
their Bose-Einstein condensation under variation of the diquark 
coupling constant $G_D$. We found that diquark molecules can appear 
at small $\mu$ and (probably realistically large) intermediate 
values of the diquark coupling, while BEC of up-down diquarks 
is realized for (probably unphysically) large values of the
diquark coupling. The dissociation temperatures of diquarks 
become higher as $G_D$ increases.
At strong coupling, the dissociation temperatures could be
higher than the critical temperature of the deconfinement 
transition.

In this work, we employed the random-phase approximation 
for the calculation of the diquark propagator.
In this approximation, the propagators in the 
polarization function Eq.~(\ref{eq:Q}) are those
for non-interacting quarks, i.e.,
the effect of diquark excitations is not self-consistently
incorporated. An extension of the present work to include
this effect is an interesting subject for further study,
because we expect that the formation of diquark molecules 
would modify the result especially at high temperatures.
Similarly, the mean-field approximation used to draw the 
phase diagram is no longer applicable due to the existence
of  well-developed soft modes in a strongly coupled system.
The incorporation of these effects has already been 
partially made in Ref.~\cite{He:2007yj}.

M.K. thanks H.~Abuki and T.~Kunihiro for discussions.


\begin{thebibliography}{99}


\bibitem{reviews} 
K.~Rajagopal and F.~Wilczek,
hep-ph/0011333;
M.~Alford,
Ann.\ Rev.\ Nucl.\ Part.\ Sci.\  {\bf 51}, 131 (2001);
T.~Sch{\"a}fer,
hep-ph/0304281;
D.~H.~Rischke, 
Prog.\ Part.\ Nucl.\ Phys.\  {\bf 52}, 197 (2004);
H.-C.~Ren,
hep-ph/0404074;
M.~Huang,
hep-ph/0409167;
I.~A.~Shovkovy,
Found.\ Phys.\ {\bf 35}, 1309 (2005) [arXiv:nucl-th/0410091].

\bibitem{KKKN02}
M.~Kitazawa, T.~Koide, T.~Kunihiro and Y.~Nemoto,
Phys.\ Rev.\ D {\bf 65}, 091504 (2002)
[arXiv:nucl-th/0111022];

\bibitem{Babaev:1999iz}
  E.~Babaev,
  Int.\ J.\ Mod.\ Phys.\  A {\bf 16}, 1175 (2001)
  [arXiv:hep-th/9909052].

\bibitem{KKKN04}
M.~Kitazawa, T.~Koide, T.~Kunihiro and Y.~Nemoto,
Phys.\ Rev.\ D {\bf 70}, 056003 (2004)
[arXiv:hep-ph/0309026];

Prog.\ Theor.\ Phys.\ {\bf 114}, 205 (2005)
[arXiv:hep-ph/0502035].

\bibitem{Mat00}
M.~Matsuzaki, Phys.\ Rev.\ D{\bf 62}, 017501 (2000).

\bibitem{AHI02}
H.~Abuki, T.~Hatsuda and K.~Itakura, 
Phys.\ Rev.\ D{\bf 65}, 074014 (2002).

\bibitem{Legg80}
A.~J.~Leggett, J.\ Phys.\ {\bf 41}, C7 (1980).

\bibitem{NSR85}
P.~Nozi\`{e}res and S.~Schmitt-Rink, J.\ Low Temp.\ Phys.\ 
{\bf 59}, 195 (1985).

\bibitem{Nishida:2005ds}
  Y.~Nishida and H.~Abuki,
  Phys.\ Rev.\  D {\bf 72}, 096004 (2005)
  [arXiv:hep-ph/0504083].

\bibitem{Nawa:2005sb}
  K.~Nawa, E.~Nakano and H.~Yabu,
  Phys.\ Rev.\  D {\bf 74}, 034017 (2006)
  [arXiv:hep-ph/0509029].

\bibitem{Abuki:2006dv}
  H.~Abuki,
  Nucl.\ Phys.\  A {\bf 791}, 117 (2007)
  [arXiv:hep-ph/0605081].

\bibitem{Deng:2006ed}
  J.~Deng, A.~Schmitt and Q.~Wang,
  arXiv:nucl-th/0611097.

\bibitem{Ebert:2006tc}
  D.~Ebert and K.~G.~Klimenko,
  Phys.\ Rev.\  D {\bf 75}, 045005 (2007)
  [arXiv:hep-ph/0611385].

\bibitem{Sun:2007fc}
  G.~f.~Sun, L.~He and P.~Zhuang,
  Phys.\ Rev.\  D {\bf 75}, 096004 (2007)
  [arXiv:hep-ph/0703159].

\bibitem{He:2007yj}
  L.~He and P.~Zhuang,
  arXiv:0705.1634 [hep-ph].

\bibitem{gSC}
  I.~Shovkovy and M.~Huang,
  Phys.\ Lett.\  B {\bf 564}, 205 (2003)
  [arXiv:hep-ph/0302142];
  M.~Alford, C.~Kouvaris and K.~Rajagopal,
  Phys.\ Rev.\ Lett.\  {\bf 92}, 222001 (2004)
  [arXiv:hep-ph/0311286];
  S.~B.~R\"uster, I.~A.~Shovkovy and D.~H.~Rischke,
  Nucl.\ Phys.\  A {\bf 743}, 127 (2004)
  [arXiv:hep-ph/0405170];
  K.~Iida, T.~Matsuura, M.~Tachibana and T.~Hatsuda,
  Phys.\ Rev.\  D {\bf 71}, 054003 (2005)
  [arXiv:hep-ph/0411356];
  S.~B.~R\"uster, V.~Werth, M.~Buballa, I.~A.~Shovkovy and D.~H.~Rischke,
  Phys.\ Rev.\  D {\bf 72}, 034004 (2005)
  [arXiv:hep-ph/0503184];
  D.~Blaschke, S.~Fredriksson, H.~Grigorian, A.~M.~\"{O}zta\c{s}, and F.~Sandin,
  Phys.\ Rev.\  D {\bf 72}, 065020 (2005)
  [arXiv:hep-ph/0503194].


\bibitem{BraunMunzinger:1999qy}
  P.~Braun-Munzinger, I.~Heppe and J.~Stachel,
  Phys.\ Lett.\  B {\bf 465}, 15 (1999)
  [arXiv:nucl-th/9903010].

\bibitem{Thou}
D.~J.~Thouless, Ann.\ Phys.\ {\bf 10}, 553 (1960).

\bibitem{Buballa:2001gj}
  M.~Buballa and M.~Oertel,
  Nucl.\ Phys.\  A {\bf 703}, 770 (2002)
  [arXiv:hep-ph/0109095].
  In this reference, since the same parameters are employed,
  the phase structure is completely the same as in 
  Fig.~\ref{fig:phased1},  
  except for the lines for $T_{\rm diss}^c$,
  which were not shown in that reference.

\bibitem{Lat_Tc}
M.~Cheng, et al., Phys.\ Rev.\ D {\bf 74}, 054507 (2006);
Y.~Aoki, et al., Phys.\ Lett\ {\bf B643}, 46 (2006).

\bibitem{Kitazawa:2002bc}
  M.~Kitazawa, T.~Koide, T.~Kunihiro and Y.~Nemoto,
  Prog.\ Theor.\ Phys.\  {\bf 108}, 929 (2002)
  [arXiv:hep-ph/0207255] and references therein.

\end{thebibliography}
\end{document}